\documentclass[english,floatfix,superscriptaddress,twocolumn,showpacs,aps,prl,print,groupedaddress]{revtex4-1}

\usepackage{amsmath}
\usepackage{graphicx}
\usepackage{bm}
\usepackage{dsfont}
\usepackage{xfrac}
\usepackage{siunitx}
\usepackage{rotating}
\usepackage{xfp}
\usepackage{booktabs}
\usepackage[resetlabels]{multibib}
\newcites{sm}{References}
\usepackage{xcolor} 
\newcommand{\vect}[1]{\boldsymbol{\mathbf{#1}}}
\newcommand{\dg}{^{\dagger }}

\usepackage{hyperref}
\hypersetup{colorlinks=true, citecolor=blue, urlcolor=blue, linkcolor=blue, breaklinks=true}

\begin{document}
\title{Algebraic Hastatic Order in One-Dimensional Two-Channel Kondo Lattice}
\author{Milan Kornja\v ca}
\author{Rebecca Flint}
\affiliation{Ames National Laboratory, U.S.~Department of Energy, Ames, Iowa 50011, USA}
\affiliation{Department of Physics and Astronomy, Iowa State University, 12 Physics Hall, Ames, Iowa 50011, USA}
\date{\today}

\begin{abstract}
The two-channel Kondo lattice likely hosts a rich array of phases, including hastatic order, a channel symmetry breaking heavy Fermi liquid.  We revisit its one-dimensional phase diagram using density matrix renormalization group and, in contrast to previous work find algebraic hastatic orders generically for stronger couplings. These are heavy Tomonaga-Luttinger liquids with nonanalyticities at Fermi vectors captured by hastatic density waves. We also find a predicted additional non-local order parameter due to interference between hastatic spinors, not present at large-N; and residual repulsive interactions at strong coupling suggesting non-Fermi-liquid physics in higher dimensions.
\end{abstract}
\maketitle

The Kondo lattice model is a foundational model of correlated electron physics, capturing how antiferromagnetic interactions between conduction electrons and local moments lead to heavy Fermi liquids, where the spins are incorporated into the Fermi surface. The model can be solved analytically in a large-$N$ limit, which captures the heavy Fermi liquid \cite{Newns1983,Coleman1983}, with $1/N$ corrections leading to magnetic order, quantum criticality and superconductivity \cite{Doniach1977,Stewart2017,Coleman2007,Qimiao2010}; numerical results in one dimension (1D) are largely consistent with large-$N$, finding a heavy Tomonaga Luttinger liquid (TLL) \cite{Tsunetsugu1993,Shibata1996,Shibata1997,Tsunegatsu1997,Xavier2002,Xavier2004,Xie2015,Basylko2008,Xie2017, Chen2023} apparently stabilized by magnetic fluctuations \cite{Khait2018}. The two-channel Kondo lattice is a well-studied extension, with two symmetry-related conduction electron channels. The impurity is quantum critical \cite{Nozieres1980,Kivelson1992,Cox1998}, with a residual Majorana fermion. The lattice has a rich interplay of spin and channel, and Majorana signatures may survive to higher dimensions \cite{Jarrell1997,Inui2020,Hu2021}. The model can be solved in two large-$N$ limits, leading to composite pair superconductivity \cite{Coleman1999,Flint2008,Hoshino2014} for $SP(N)$ and channel symmetry breaking heavy Fermi liquids known as hastatic order for $SU(N)$ \cite{Chandra2013,Hoshino2011,Hoshino2012,Hoshino2013,Zhang2018,Wugalter2019, Ge2022, Ge2024}. The physical $N=2$ limit has been studied numerically in 1D \cite{Moreno2001,Shauerte2005} and infinite dimensions  \cite{Jarrell1996,Jarrell1997,Nourafkan2008,Hoshino2011,Hoshino2012,Hoshino2013,Hoshino2014}, but many questions remain, including the general phase diagram; validity of large-$N$ limits, including the presence of composite pair superconductivity; and the nature and Fermi wave-vectors of metallic phases, including non-Fermi liquid signatures. 

\begin{figure}[!htb]
\vspace*{-0.0cm}
\includegraphics[width=1.0\columnwidth]{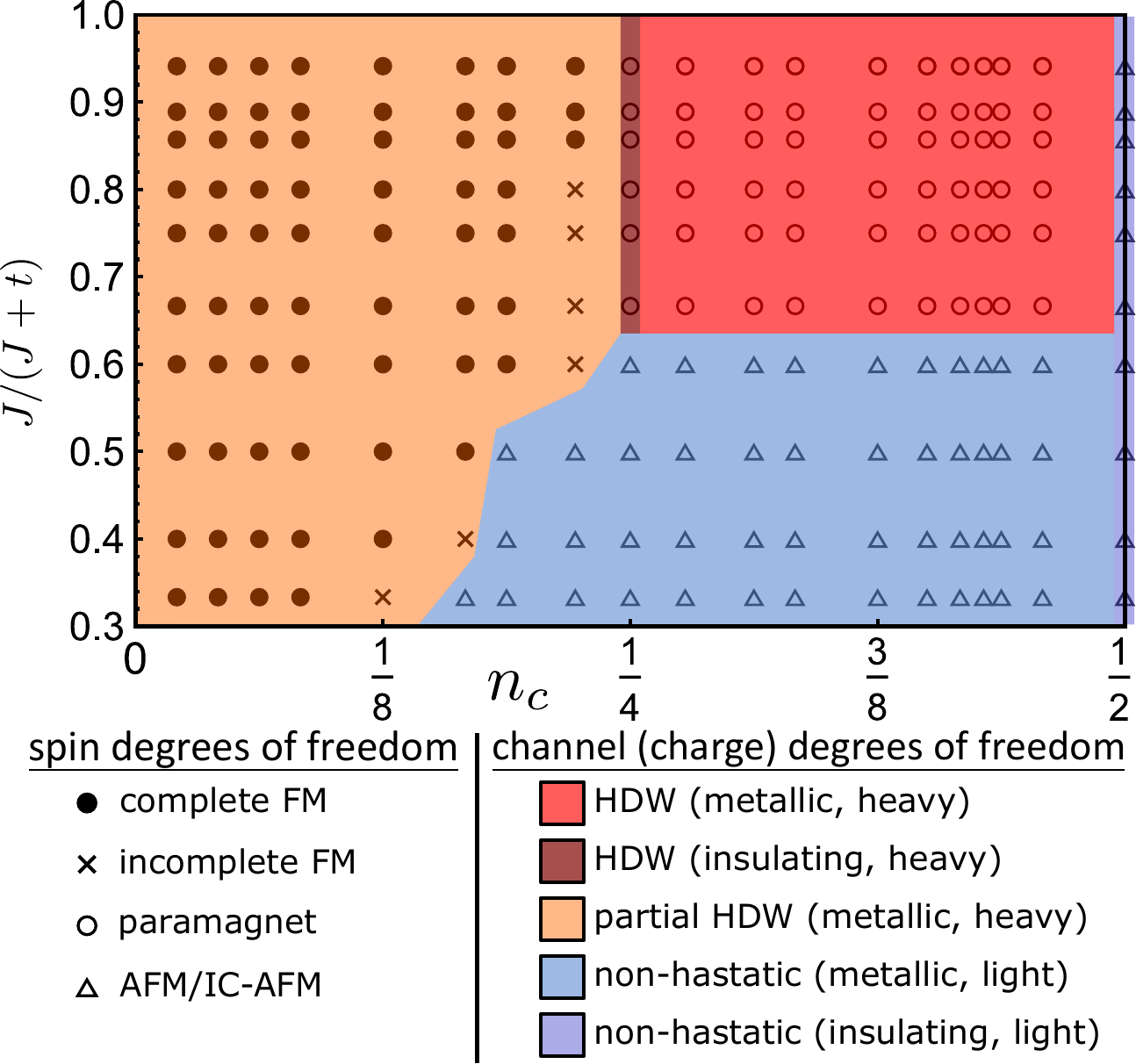}%
\vskip -0.1cm
\caption{Zero temperature phase diagram of the 1D two-channel Kondo lattice obtained with DMRG, as a function of conduction electron filling ($n_c$) and Kondo coupling ($J$). The ground state has magnetic and channel/hastatic orders, denoted by symbols and colors as described in the legend. 
Regions are ``heavy'', incorporating at least some spins into the Fermi sea, or ``light''. Quarter- and half-filling are commensurate hastatic density wave (HDW) and antiferromagnetic (AFM) insulators, respectively, while all other regions are metallic with generically incommensurate orders. There are three distinct metallic regions: a coexisting ferromagnet and HDW (FM-HDW) for $n_c < 1/4$, where $n_c$ spins are screened and the rest order ferromagnetically; a pure HDW for stronger coupling and $1/4 < n_c < 1/2$; and a weak-coupling AFM without hastatic correlations (IC-AFM).
\label{fig:Fig1PD}}
\vskip -0.3cm
\end{figure}

Given recent analytical insights \cite{Chandra2013,Zhang2018, Wugalter2019,Kornjaca2020} and computational improvements, we revisit the 1D two-channel Kondo lattice using density matrix renormalization group (DMRG). We can now address both potential order parameters and Luttinger liquid properties, including whether there are ``heavy'' TLLs.  Our results are summarized in Fig. \ref{fig:Fig1PD}. We find algebraic hastatic order at larger couplings for all conduction electron fillings except half-filling, and we find these are heavy TLLs incorporating the spins into the Fermi surface.  Surprisingly, the two-channel combination of channel and magnetic fluctuations appears to be more effective in stabilizing the heavy TLL than the one-channel model \cite{Khait2018}.  Our study greatly extends previous results \cite{Shauerte2005}, which found algebraic hastatic order only at quarter-filling and did not examine the nature of the metallic phases.  Our results have implications for non-Kramers doublet materials, where two-channel Kondo physics may be relevant for unconventional superconductivity in UBe$_{13}$ \cite{Cox1987,Cox1996,Jarrell1996,Jarrell1997}, 1-2-20 Pr-based materials physics \cite{Yoshida2017,Iwasa2017,Zhang2018,VanDyke2019,Inui2020}, and hidden order in URu$_2$Si$_2$ \cite{Chandra2013, Chandra2015, Kornjaca2020}. 

The 1D two-channel Kondo lattice model is:
\begin{equation}\label{eq:Ham}
    H\!=\! -t\!\sum_{i\alpha\sigma}\!c\dg_{i\alpha\sigma}c_{i+1\alpha\sigma}+H.c.+\frac{J}{2}\!\!\!\sum_{i\alpha\sigma\sigma'}\!\!\vect{S}_{fi}\!\cdot \!c\dg_{i\alpha\sigma}\vect{\sigma}_{\sigma\sigma'} c_{i\alpha\sigma'}
\end{equation}
where $t=1$ is the conduction electron hopping and $J >0$ is the Kondo coupling. $i$ labels sites ($1 \leq i\leq L$), where each site has both conduction electrons ($c^\dagger_{i\alpha \sigma}$) with spin $\sigma$ and channel $\alpha$, and local $S = 1/2$ moments ($\vect{S}_{fi}$). 
$\vect{\sigma}$ are the Pauli matrices in spin space. We fix the conduction electron filling, $0 \leq n_c \leq 1$, with $n_c = 1$ indicating four electrons per site. This is the simplest Kondo model with both $SU(2)$ spin and $SU(2)$ channel symmetries.

The channel degeneracy leads to a rich possibility of phases.  These can be divided into ``heavy'' and ``light'' phases, where the spins either are incorporated into the Fermi sea or remain decoupled; in the light phases these typically order magnetically, $\langle \vect{S}_{fi}\rangle \neq 0$.  There are two proposed ``heavy'' orders: a channel symmetry breaking heavy Fermi liquid that we call hastatic order \cite{Hoshino2011,Chandra2013}, and composite pair superconductivity \cite{Coleman1999}, which incorporates the spins directly into heavy Cooper pairs.  These orders arise within different large-$N$ limits of the $SU(2)$ two-channel model, where the more commonly used $SU(N)$ limit leads to hastatic order, and composite pairing arises in the symplectic-$N$ limit extending $SU(2)$ to $SP(N)$ \cite{Flint2008}.  For $N=2$ and $n_c=1/2$, these are components of an $SO(5)$ composite order parameter.  Both have been found in $d=\infty$ away from half-filling \cite{Hoshino2011,Hoshino2012,Hoshino2013,Hoshino2014}, and both form either uniform or modulated orders.

The hastatic order parameter is a composite order parameter of conduction and $f$-moments, $\vec{\Psi}$ that captures the Kondo singlet channel polarization:
\begin{equation}\label{eq:PsiOP}
    \vect{\Psi}(i)=\frac{1}{2}\sum_{\sigma\sigma'\alpha\alpha'}c\dg_{i\alpha\sigma}\vect{\sigma}_{\sigma\sigma'}\vect{\tau}_{\alpha\alpha'}c_{i\alpha'\sigma'}\cdot \vect{S}_{fi},
\end{equation}
where $\vect{\tau}$ are channel Pauli matrices. The $\hat z$ component is manifestly   channel polarization, $\Psi^{z}(i)=(\vect{S}_{ci,\alpha=+}-\vect{S}_{ci,\alpha=-})\cdot \vect{S}_{fi}$ \cite{Hoshino2011}. Hastatic order also generates staggered channel polarization in the conduction electrons,  $n_{i\alpha\sigma}=c\dg_{i\alpha\sigma}c_{i\alpha\sigma}$, whose correlations were used previously \cite{Shauerte2005}.

The complex composite pair order parameter is,
\begin{equation}\label{eq:CPOP}
    \Delta_{CP}(j)=\sum_{\alpha\sigma\sigma'}c\dg_{j\alpha\sigma'}\left[\vect{\sigma} (i\sigma_2)\right]_{\sigma\sigma'}c\dg_{j\bar{\alpha}\sigma'}\cdot \vect{S}_{fj},
\end{equation}
with $\bar{\alpha}=-\alpha$.  Electrons in orthogonal channels screen the same spin, giving singlet superconductivity.

An additional hastatic order parameter was recently predicted \cite{Kornjaca2020}.  In large-$N$, hastatic order has a \emph{spinorial} order parameter, $\langle V_{i\alpha}\rangle$ representing a channel-dependent Kondo singlet.  This quantity is gauge dependent, meaning finite-$N$ order parameters must be bilinears.  The composite order parameter, $\vect{\Psi}(i) \propto \sum_{\alpha\alpha'}\langle V_{i\alpha}^*\vect{\tau}_{\alpha\alpha'} V_{i\alpha'}\rangle$, is associated with the on-site moments of these spinors, but another, non-local order parameter can arise from interference between spinors at different sites, breaking additional symmetries \cite{Zhang2018,Kornjaca2020}.  This interference requires intersite spin correlations, and is not present for $N=\infty$.  $1/N^2$ RKKY couplings generate these correlations \cite{Houghton1987,Andrei1989}, often treated within large-$N$ Kondo-Heisenberg models \cite{Andrei1989,Zhang2018} as an emergent f-hopping, $t_{f,ij}$ describing spin-liquid physics. This order parameter, $\vec{\Phi}(i,j) \propto \sum_{\alpha\alpha'}\langle t_{f,ij} V_{i\alpha}^*\vect{\tau}_{\alpha\alpha'} V_{j\alpha'}\rangle$ is only present in modulated hastatic phases, and may be written as \cite{Kornjaca2020}: 
\begin{equation}\label{eq:Phidef}
    \Phi^a(i,j)
    = i \sum_{\sigma\sigma'\alpha\alpha'} \vect{S}_{f,i} \cdot \left( c\dg_{i\alpha\sigma} \tau^a_{\alpha \alpha'}\vect{\sigma}_{\sigma \sigma'}c_{j\alpha'\sigma'}\times \vect{S}_{f,j}\right),
\end{equation}
where $i,j$ denote the sites, typically nearest-neighbors, whose spinorial interference generates $\vec{\Phi}$. The coexistence of a local, composite order parameter $\vec{\Psi}$ and a non-local, spin-liquid-like order parameter, $\vec{\Phi}$ is reminiscent of the order parameter fragmentation found in spin ice \cite{Brooks2014,Petit2016,Erten2023}, and is an explicit example of Kondo order parameter fractionalization \cite{Komijani2018, Tsvelik2022}.

We obtained the ground state phase diagram using finite system DMRG \cite{White1992,White1993} in an ITensor implementation \cite{itensor} with open boundary conditions.  We conserve the total conduction electron number, $n_c=\frac{1}{4L}\sum_{i\sigma\alpha}n_{i\alpha\sigma}$,
 and the $z$ component of total angular momentum, $S^{z}=\sum_{i}[S_{fi}^z+\sum_{\alpha}S_{ci\alpha}^z]$, where $\vect{S}_{ci\alpha}=\frac{1}{2}\sum_{\sigma\sigma'}c\dg_{i\alpha\sigma}\vect{\sigma}_{\sigma\sigma'}c_{i\alpha\sigma'}$ is the conduction electron spin for a given site and channel. We use bond dimensions of up to $m=5000$ on lattices of up to $L=96$ sites, resulting in a maximum discarded weight of $10^{-6}$ (with $<10^{-8}$ typical in the strong coupling regions), implying generally good convergence. All figures used $L = 72$. The phase diagram shown in Fig. \ref{fig:Fig1PD} was mapped out via ground state correlation functions in the lowest energy total spin sectors as a function of $n_c$, for $0 \leq n_c \leq 1/2$, and $J$ for $1/3 \leq J/(J+t)\leq 16/17$. 

We examined the following correlation functions,
\begin{align}\label{eq:SPsix}
    S_{\Psi}(x) &=\langle \Psi^{z}(i) \Psi^{z}(i+x) \rangle \cr
    S_{\Phi}(x)&=\langle  \Phi^z(i,i+1)  \Phi^{z\dagger}(i+x,i+1+x)\rangle\cr
    S_{f}(x) &=\langle S_{fi}^{z} S_{f,i+x}^{z}\rangle.
\end{align}
The first measures composite order correlations, which captures the same physics as the conduction channel polarization correlations, $D(x) =\sum_{\sigma\sigma'\alpha\alpha'}\alpha\alpha'\langle n_{i\alpha\sigma}  n_{i+x\alpha'\sigma'}\rangle$ used previously  \cite{Shauerte2005}.  $S_{\Phi}(x)$ captures correlations of the nearest-neighbor $\Phi$ order parameter, while $S_f(x)$ captures spin correlations. We fix the reference site $i=10$, but results are unchanged by averaging over $i$'s sufficiently far from the edges; $x$ is a discrete variable representing distances between sites.  We use the $SU(2)$ symmetries to fix the spin/channel components to $\hat z$.

\begin{figure}[!tb]
\vspace*{-0cm}
\includegraphics[width=1.0\columnwidth]{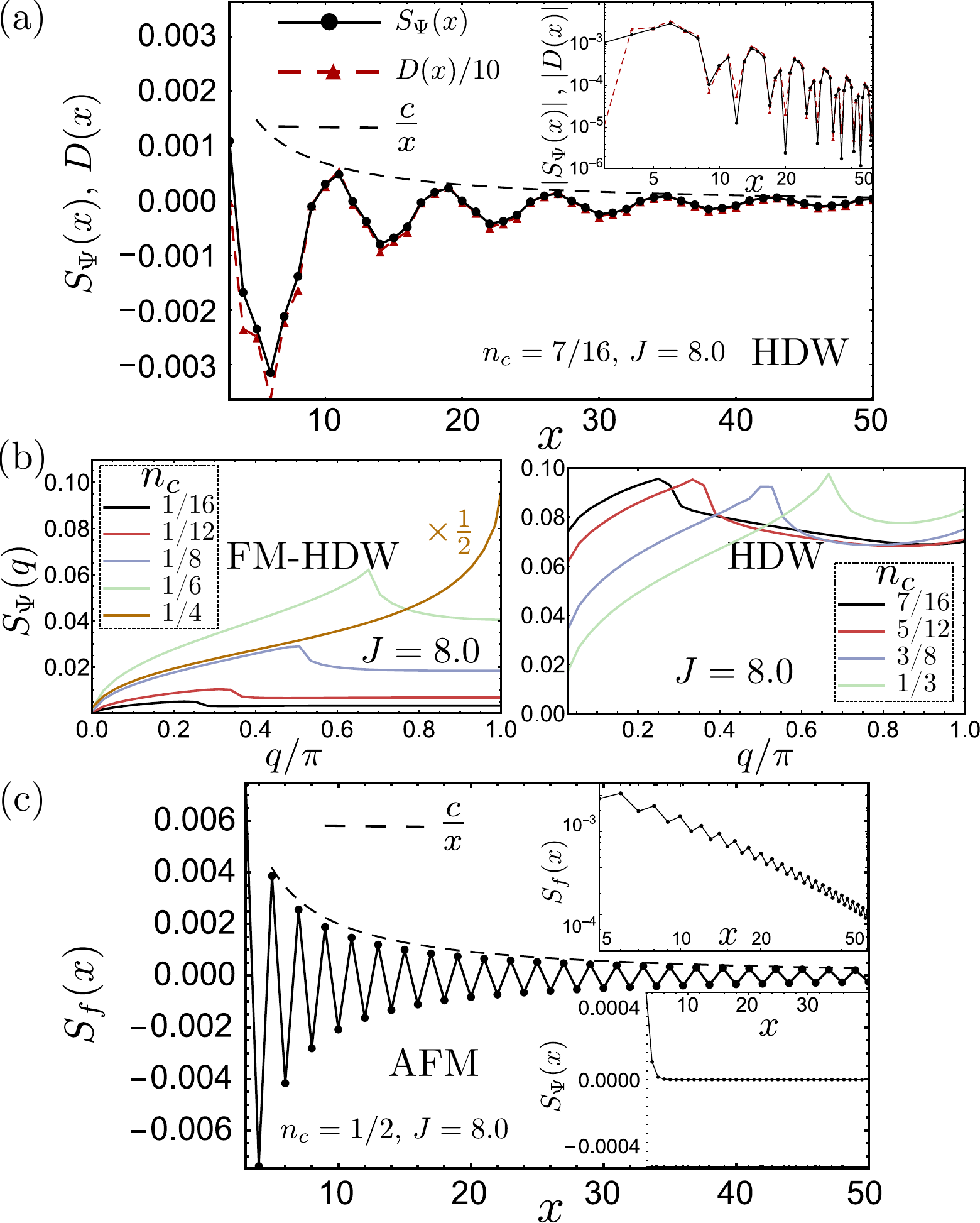}%
\vskip -0.0cm
\caption{(a)  Example hastatic  correlator, $S_\Psi(x)$, with $n_c = 7/16$, $J = 8.0$. It decays algebraically, with a clear modulation. Note, $D(x)$ \cite{Shauerte2005} closely follows $S_\Psi(x)$, up to a numerical factor. The inset log-log plot confirms the power law decay. $S_\Psi(x)$ is similar for the FM-HDW and HDW.
(b) The Fourier transform, $S_\Psi(q)$ for several $n_c$ with $J = 8.0$. The $Q$ obtained from the peak location smoothly evolves from nearly zero at low fillings to $\pi$ at $1/4$ filling as $Q=4\pi n_c$ in the FM-HDW (left); the weight under $S_\Psi(q) \propto n_c^2$, with the $n_c = 1/4$ curve reduced by $1/2$. In the HDW ($n_c > 1/4$) (right), $Q$ evolves smoothly from $\pi$ as $Q=4\pi (1/2-n_c)$. (c)  In the AFM insulator ($n_c = 1/2$), the spin correlator, $S_f(x)$ has a $1/x$ power law (log-log plot in top inset), with $Q = \pi$, while $S_\Psi(x)$ (bottom inset) decays exponentially.
\label{fig:Fig2psicor}}
\end{figure}

The peculiarities of 1D allow DMRG to efficiently study this model, but also mean there is no long range order, only algebraic or exponential correlations, and Luttinger instead of Fermi liquids. The nature of these gapless phases is indicated by their central charge, which can be calculated via the scaling of entanglement entropy with  system size cut \cite{Vidal2003, Laflorencie2006}, as shown in the supplementary information \cite{SM, Hui-Ke2023, Andrei2000,  Voit1995, coleman2015book}  We find that all metallic HDWs have central charge, $c=2$, while the insulating quarter-filled HDW has $c=1$, as does the insulating half-filled AFM. Integer central charge implies that all HDW regions are Tomonaga-Luttinger liquids (TLLs), although how the charge, spin and channel sectors contribute in the metallic HDWs is an open question. While TLLs do not have a jump at the Fermi wave-vector, they have nonanalyticities at $k_F$ that manifest in both spin and charge Friedel oscillations, whose Fourier transforms have peaks at $2k_F$ and $4k_F$ \cite{Shibata1996,Shibata1997,Khait2018}, as well as directly in the conduction electron momentum distribution \cite{Voit1995, Eidelstein2011, Xie2015}, 
\begin{equation}\label{eq:nq}
    n_q=\frac{1}{L}\sum_{ij\alpha\sigma}e^{iq(i-j)}\langle c\dg_{i\alpha\sigma}c_{j\alpha\sigma}\rangle.
\end{equation}
We sum over all $i,j$, but results are similar if sites near the edges are excluded, suggesting that our results represent bulk physics.
We can now show that the HDW Fermi surfaces incorporate the spins \cite{coleman2015book, Oshikawa2000, Hazra2021,Shibata1996,Shibata1997,Xavier2002,Xavier2004,Xie2015,Xie2017,Khait2018}, confirming a key large-$N$ two-channel result \cite{Chandra2013, Zhang2018, Wugalter2019, Ge2022}.  We will show only $n_q$, which detects $k_F$ directly, but the $2k_F$ and $4k_F$ Friedel oscillation peaks were used as checks \cite{SM}.  The relative weight of these peaks and the overall spatial dependence of the Friedel oscillations can be used to extract the charge Luttinger parameter, $K$, which contains information about residual interactions \cite{Voit1995,Shibata1997,Khait2018}. Note that TLL have intrinsic algebraic spin, charge and superconducting correlations, all with $1/x^\alpha$, $\alpha > 1$ power laws, while hastatic correlations, when present, dominate with $\alpha = 1$. 

\begin{figure}[!tb]
\vspace*{-0cm}
\includegraphics[width=1.0\columnwidth]{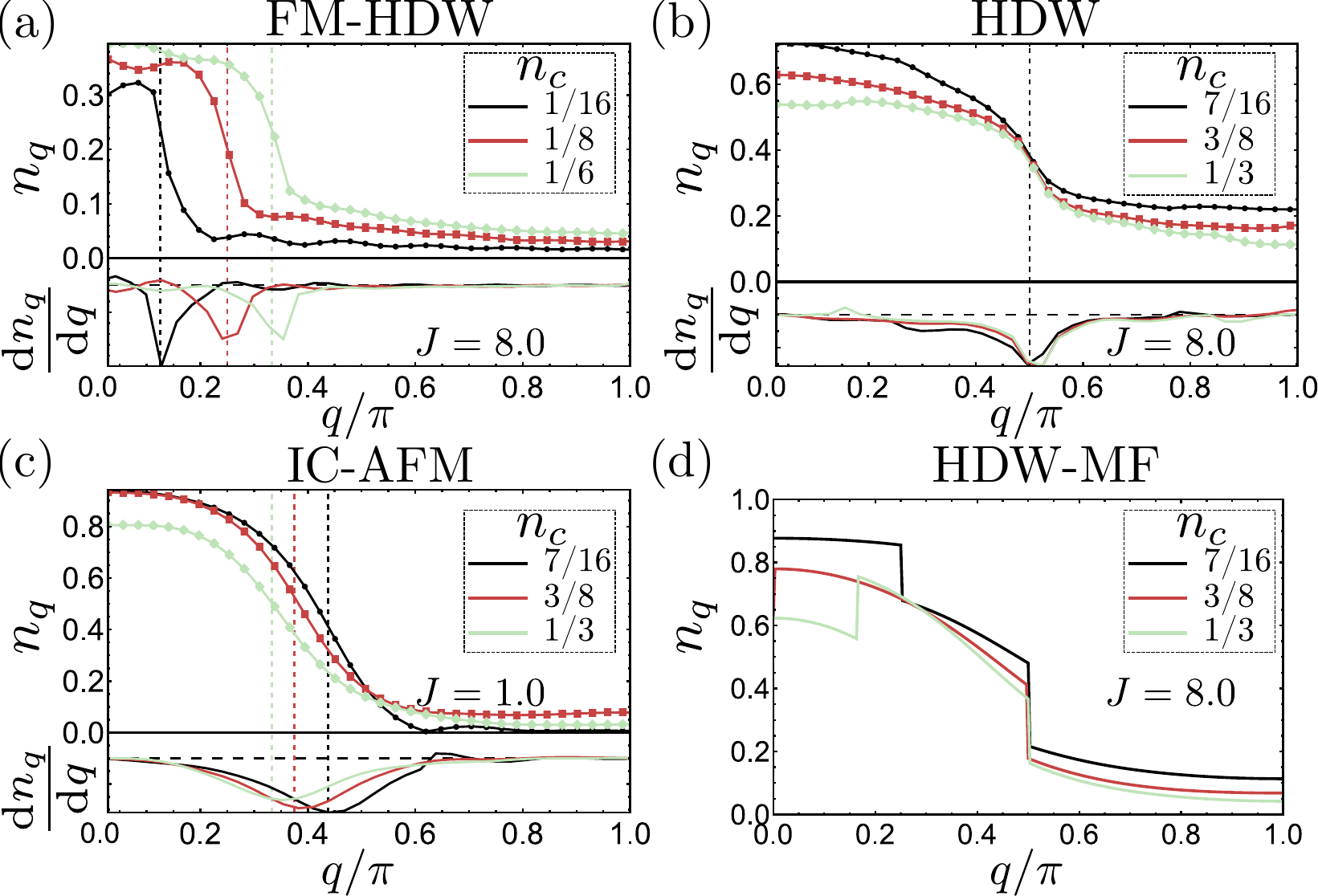}%
\vskip -0.0cm
\caption{For several $n_c$ within each metallic region, we calculate the conduction electron momentum distribution, $n_q$, which should have nonanalyticities at the Fermi wave-vectors, with the derivative ($dn_q/dq$, arbitrary units) plotted below.  The numerical results are in (a-c). (a) shows the FM-HDW $n_q$, with dashed lines indicating the nonanalyticity at $k_F^*=2 \pi n_c$, twice the light $k_F$, consistent with incorporating $n_c$ spins into the Fermi sea. (b) shows the HDW $n_q$, where the primary $k_F^* = \pi/2$, regardless of filling, with secondary $n_c$ dependent nonanalyticities at lower $q$. (c) shows $n_q$ in the non-hastatic IC-AFM, with light $k_F = \pi n_c$. In (d), we calculate $n_q$ for the HDW within a large-$N$ mean-field theory \cite{SM}. Only some heavy Fermi surfaces have non-zero \emph{conduction} weight ($n_q$), including one pinned at $\pi/2$; the secondary nonanalyticity at low q is also qualitatively captured. \label{fig:Fig3FS}}
\end{figure}

Now we turn to the nature of the five distinct ground states found in the phase diagram in Fig. \ref{fig:Fig1PD}.  We confirm the spin correlations previously reported \cite{Shauerte2005}, but find additional hastatic correlations away from $n_c=1/4$.

For low filling ($n_c < 1/4$) and moderate to strong coupling, there is a ferromagnetic region with coexisting algebraic hastatic order (FM-HDW).  $n_c$ spins are screened by forming Kondo singlets, while the remaining spins are fully polarized with $S = S_{max}=(1-4n_c)L/2$, analogous to the single-channel case \cite{Tsunegatsu1997,Peters2012}, aside from a small incomplete ferromagnetic region near the phase boundary ($0 < S\leq S_{max}$).  Fig. \ref{fig:Fig2psicor}(b) shows the $\Psi$ structure factor, $S_{\Psi}(q)=\frac{1}{L}\sum_{x}S_{\Psi}(x)e^{-iqx}$.  The peak position gives the HDW $Q$-vector, $Q = 4\pi n_c$, which approaches $\pi$ at $1/4$-filling. $S_\Psi(x)$ decays algebraically as $1/x$, while the spin correlations are those of a TLL.  In Fig. \ref{fig:Fig3FS}(a), $n_q$ has a nonanalyticity at $k_F^* = 2 \pi n_c$, twice the light $k_F = \pi n_c$, indicating that $n_c$ spins are incorporated into the Fermi surface. The total weight under $S_\Psi(q)$ depends on the number of screened spins, growing as $n_c^2$. While the FM-HDW is a TLL, the charge Luttinger parameter $K(J)$ decreases with $J$, suggesting increasingly repulsive residual interactions as strong coupling is approached \cite{SM}. 

For $1/4 \leq n_c < 1/2$ and $J/(J+t) \geq 3/5$, there is a purely hastatic region (HDW), with fully screened spins ($S = 0$) and algebraic hastatic order at $Q = 4\pi(1/2-n_c)$, with correlations again decaying as $1/x$.  Exactly at $n_c = 1/4$, this is a hastatic Kondo insulator, with $Q = \pi$, and charge and spin gaps. Elsewhere, it is metallic - a heavy TLL, as seen in $n_q$ in Fig. \ref{fig:Fig3FS}(b).  The main nonanalyticity of $n_q$ is pinned to $k_F^*=\pi/2$, regardless of filling.  This pinning results from the $Q$ dependence of $n_c$, and is captured within a large-$N$ HDW mean-field theory \cite{SM}.  The mean-field heavy bands are obtained at commensurate $n_c$, and the resulting conduction electron momentum distribution, $n_q$ is shown in Fig. \ref{fig:Fig3FS}(d).  While there are many \emph{heavy} band crossings, due to the large unit cell, we find typically only one or two Fermi surface jumps in $n_q$: one pinned to $k_F = \pi/2$ and an $n_c$ dependent one at lower $q$.  Both are consistent with DMRG results in location and sign [compare Fig. \ref{fig:Fig3FS}(b),(d)].

At $1/2$-filling, we find an antiferromagnetic insulator (AFM), whose spin correlations decay as $1/x$ [Fig. \ref{fig:Fig2psicor}(c)], with no spin gap. The hastatic correlations decay rapidly, suggesting the spins and conduction electrons are decoupled, although the electrons also have staggered algebraic spin correlations.

For weaker coupling, there is a non-hastatic, incommensurate antiferromagnet (IC-AFM), where spins and conduction electrons have $Q = 2k_F = 2\pi n_c$, consistent with a spin-density wave of the light Fermi surface.  $n_q$ also shows this light $k_F=\pi n_c$, as seen by peaks in $dn_q/dq$ in Fig. \ref{fig:Fig3FS}(c).  Our results here are less reliable, as weak coupling  ($J\lesssim 1.5$) is inherently harder to treat with DMRG, manifesting as higher maximum discarded weights. We therefore cannot conclusively calculate the central charge to confirm the proposed fractional central charge in the weak Kondo limit of the two-channel Kondo-Heisenberg model \cite{Andrei2000, Ge2022}. This region appears to have no charge or spin gaps and is reminiscent of the RKKY liquid phase hypothesized for $J \ll t$ in the single-channel lattice \cite{Xavier2002, Shauerte2005, Schimmel2016, Khait2018}.

\begin{figure}[!htb]
\vspace*{-0cm}
\includegraphics[width=1.0\columnwidth]{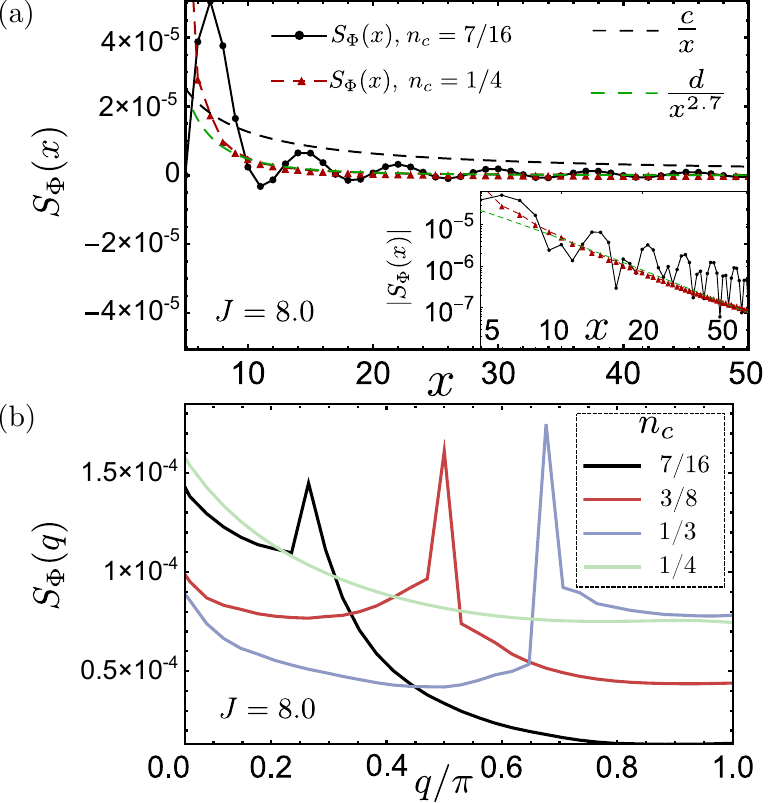}%
\vskip -0.0cm
\caption{The non-local order parameter, $\vec{\Phi}$ captures interference between neighboring hastatic spinors. (a) $S_\Phi(x)$ correlator at $J=8.0$, $n_c = 1/4$ (insulating HDW) and $7/16$ (HDW). There is a clear power law dependence, with a smaller magnitude than $S_\Psi$. In metallic regions, $S_\Phi$ shows $1/x$ behavior with a clear oscillation. In the insulator, $S_\Phi$ is uniform, with a sub-leading $1/x^\alpha$ power law \cite{Kornjaca2020}; we fit $\alpha = 2.7$ for $x \geq 20$. $\vec{\Phi}$ is not present for large-$N$ and indicates RKKY physics. (inset) Log-log plot confirming power-law decay; $n_c = 1/4$ has a steeper slope than $n_c = 7/16$, due to differing exponents. (b) The Fourier transform, $S_\Phi(q)$ for several $n_c$, $J=8.0$. $n_c = 1/4$ only has a $Q = 0$ peak, but other fillings show both uniform and oscillating components, with the same $Q$ as $S_\Psi$. \label{fig:Fig4phicor}}
\end{figure}

We also examined composite pair correlations, $S_{CP}(x)=\langle \Delta_{CP}(i) \Delta\dg_{CP}(i+x)\rangle$, finding exponential suppression with generically short correlation lengths, by contrast to the algebraic TLL conventional superconducting correlations. In the HDW, $\Delta_{CP}(i)$ is staggered, and the correlation length increases with $J$ and $n_c$  \cite{SM}, consistent with the order found for $d=\infty$ \cite{Hoshino2014}. 

Finally, we examine the additional hastatic order parameter, $\vec{\Phi}$, which captures interference between neighboring hastatic spinors.  $\vec{\Phi}$ has algebraic correlations in the HDWs, which must arise from RKKY interactions beyond large-$N$. The correlations shown in Fig. \ref{fig:Fig4phicor} are consistent with simple Landau arguments \cite{Kornjaca2020} predicting uniform and $Q$-modulated components of $\vec{\Phi}$, given a modulated $\vec{\Psi}(Q)$.  Away from quarter filling, $S_\Phi$ has $1/x$ correlations, with the leading $Q$-vector matching $\vec{\Psi}$, and a potential uniform component. At quarter filling, $\vec{\Phi}$ is uniform, with a distinct sub-leading $1/x^\alpha$ power law; $\alpha \approx 3$, but is sensitive to the $x$ fitting range, with $\alpha \in \{2.5, 3.5\}$.

To conclude, we find hastatic correlations to be nearly ubiquitous for strong coupling, and have shown that these regions are heavy TLL whose Fermi surfaces can be understood within large-$N$.
The two-channel Kondo insulator at $n_c = 1/4$ is particularly robust, which is relevant for its predicted Majorana zero modes \cite{Majorana2021}. We also find a predicted hastatic order parameter associated with inter-site spinorial interference, which implicates RKKY physics, and signatures of a residual critical nature of the TLL: in the FM-HDW, the charge Luttinger parameter, $K$ decreases as $J$ increases, meaning the residual interactions are \emph{increasingly} repulsive approaching strong coupling \cite{SM}, opposite to the single-channel case \cite{Shibata1997,Khait2018}. Further work is needed to resolve $K(J)$ in the HDW and to address whether higher dimensional hastatic phases are non-Fermi liquids \cite{Jarrell1997,Inui2020,Hu2021}.

\begin{acknowledgments}
We acknowledge stimulating discussions with Bryan Clark, Eduardo Miranda, and Victor L. Quito. This work was primarily supported by the U.S. Department of Energy, Office of Science, Basic Energy Sciences, under Award No. DE-SC001589; central charge calculations were added later and supported by the U.S. DOE,  Basic Energy Sciences, Materials Science and Engineering. 
\end{acknowledgments}



%


\clearpage
\onecolumngrid
\begin{center}
{\bf Supplemental Material: ``Algebraic Hastatic Order in One-Dimensional Two-Channel Kondo Lattice''}
\end{center}
\vspace{.5cm}%

In this supplemental material, we provide the details of calculations used in the main text, and some additional DMRG data supporting our conclusions. We start by presenting the large-N mean-field theory of the pure HDW. After that, we show that our critical HDW phases correspond to Tomonaga-Luttinger liquids and support our Fermi surface results with an analysis of the Friedel oscillations. Lastly, we explore the composite pair correlations. We also present several examples establishing the satisfactory convergence of our simulations.

\tableofcontents

\section{I. Mean-field theory of the HDW}

In this section we develop the large-$N$ mean-field theory of the pure hastatic density wave (HDW) region that is found for $1/4 \leq n_c < 1/2$ and strong coupling, and show the self-consistent mean-field band-structures. To facilitate the large-$N$ expansion, the spin degree of freedom is promoted from $SU(2)$ to $SU(N)$ with $\vect{S}_{fi}=\frac{1}{2}\sum_{\sigma\sigma'}f\dg_{i\sigma}\vect{\sigma}_{\sigma\sigma'} f_{i\sigma'}$, where $\vect{\sigma}$ now represents the $SU(N)$ generator (and similarly for conduction electron spins) \cite{Newns1983,Coleman1983}. This representation is faithful if $f$-fermions are exactly half-filled on each site, which can be enforced by a local constraint field, $\lambda_i$.  The $SU(N)$ two-channel Kondo Hamiltonian is then: 

\begin{align}\label{eq:hamff}
    H=&-t\sum_{i\alpha\sigma}\left(c\dg_{i\alpha\sigma}c_{i+1,\alpha\sigma}+h.c.\right)  +\frac{J}{N}\sum_{i\alpha\sigma\sigma'}f\dg_{i\sigma}c_{i\alpha\sigma}c\dg_{i\alpha\sigma'}f_{i\sigma'}\\
    &+\sum_{i}\lambda_i\left(\sum_{\sigma}f\dg_{i\sigma}f_{i\sigma}-\frac{N}{2}\right)-\mu\sum_i\left(\sum_{\alpha\sigma}c\dg_{i\alpha\sigma}c_{i\alpha\sigma}-N n_c\right).\notag
\end{align}
Here, we switched to the grand canonical ensemble by introducing a chemical potential for the conduction electrons, $\mu$. The resulting Hamiltonian has quartic interactions and can be Hubbard-Stratonovich decoupled using a channel dependent hybridization \cite{Zhang2018,Wugalter2019, Ge2022}:
\begin{equation}\label{eq:Vdef}
    V_{i\alpha}=\frac{J}{N}\sum_{\sigma}\langle f\dg_{i\sigma}c_{i\alpha\sigma}\rangle,
\end{equation}
to yield the Hamiltonian:
\begin{align}\label{eq:hamMF}
    H_{MF}=&-t\sum_{i\alpha\sigma}\left(c\dg_{i\alpha\sigma}c_{i+1,\alpha\sigma}+h.c.\right)  +\sum_{i\alpha\sigma'}\left(V_{i\alpha}c\dg_{i\alpha\sigma'}f_{i\sigma'}+h.c.\right)+\sum_{i}\lambda_i\left(\sum_{\sigma}f\dg_{i\sigma}f_{i\sigma}-\frac{N}{2}\right)\\
    &-\mu\sum_i\left(\sum_{\alpha\sigma}c\dg_{i\alpha\sigma}c_{i\alpha\sigma}-N n_c\right)+\sum_{i\alpha}\frac{N \left|V_{i\alpha}\right|^2}{J},\notag.
\end{align}
Mean-field values for the $V_{i\alpha}$ and $\lambda_i$ can be obtained by taking the saddle point approximation in a path integral approach, which is exact as $N\rightarrow\infty$. The mean-field results are expected to apply well to $N=2$ in higher dimensions \cite{Coleman1983, Chandra2013, Zhang2018,Wugalter2019}.

We now make a HDW Ansatz for the mean-fields, with the $Q$ found in our DMRG results, and assuming uniform $|V|$ and $\lambda_i=\lambda$, which satisfies the constraint on average. The two-component spinorial hybridization is defined by its (uniform) magnitude and two spatially varying angles, $\theta_i$ and $\phi_i$:
\begin{equation}\label{eq:VQ}
\vect{V}_i = |V|\left( \begin{array}{c} \cos \frac{\theta_i}{2} \mathrm{e}^{i\phi_i/2} \\ \sin \frac{\theta_i}{2} \mathrm{e}^{-i\phi_i/2}\end{array}\right).
\end{equation}
In the simple mean-field above, we set $\phi_i = 0$. $\theta_i$ is then modulated consistent with the HDW $Q =2\pi(1-2n_c)$, which requires enlarging the unit cell by a factor  $M=\frac{\kappa}{1-2n_c}$, where $\kappa$ is the smallest integer such that $M$ is an integer. $\phi_i = 0$ is not the most general HDW Ansatz for a given $\mathbf{Q}$, as $\phi_i$ modulation is also possible, which give a helical Ansatz.  We find, however, that our planar Ansatz is already successful in qualitatively capturing the HDW wave-vectors. We proceed to define $\theta_i= \gamma[i]\frac{2\pi}{M} $, where $\gamma [i]=i\pmod{M}$ is an effective basis index within the new unit cell.

The mean-field Hamiltonian is quadratic and can be readily solved for our HDW Ansatz. To facilitate this, we first Fourier transform the conduction electrons and the $f$-fermions in the reduced Brillouin zone (BZ) appropriate for HDW Q Ansatz:
\begin{equation}\label{eq:FTmcell}
    c_{q\gamma\alpha\sigma}=\sqrt{\frac{M}{L}}\sum_{j}e^{-iqj}c_{j\alpha\sigma} \delta\left[j\!\!\!\!\!\pmod{M}-\gamma\right], \qquad  f_{q\gamma\sigma}=\sqrt{\frac{M}{L}}\sum_{j}e^{-iqj}f_{j\sigma}\delta\left[j\!\!\!\!\!\pmod{M}-\gamma\right].
\end{equation}
Here $q$ denotes the momentum in the reduced BZ, while $\gamma$ is the basis index within the enlarged unit cell, enforced by the delta functions. The resulting Hamiltonian can be diagonalized by the unitary transformation:
\begin{equation}\label{eq:betadiag}
    \beta_{q\eta\sigma}=\sum_{\gamma}\left(\sum_{\alpha}U_{\eta,\gamma\alpha}(q)c_{q\gamma\alpha\sigma}+V_{\eta,\gamma}(q)f_{q\gamma\sigma}\right),
\end{equation}
where $U$ and $V$ are parts of the unitary transformation matrix $T=[U\, \, V]$ such that the resulting Hamiltonian is diagonal:
\begin{equation}\label{eq:hamdiag}
    H_{MF}=\sum_{q\eta\sigma}E_{q\eta}\beta\dg_{q\eta\sigma}\beta_{q\eta\sigma},
\end{equation}
with hybridized band dispersion $E_{q\eta}$ for $\eta=1,..., 3M$ bands.

We then solve for the mean-field parameters $|V|$, $\lambda$ and $\mu$ self-consistently by minimizing the ground state energy,
\begin{equation}\label{eq:Fmf}
    \mathcal{F}_{MF}=\sum_{q\eta\sigma}E_{q\eta}\theta(-E_{q\eta}), \qquad     \left(\frac{\partial  \mathcal{F}_{MF}}{\partial |V|},\frac{\partial  \mathcal{F}_{MF}}{\partial \lambda},\frac{\partial  \mathcal{F}_{MF}}{\partial \mu}\right)=0,
\end{equation}
for fixed $n_c$ and $J $.  

\begin{figure}[!htb]
\vspace*{-0cm}
\includegraphics[width=0.95\columnwidth]{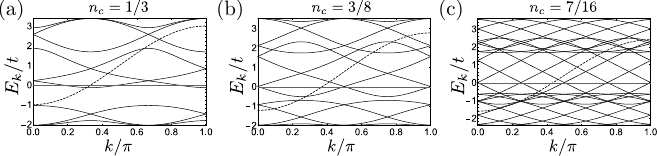}%
\vskip -0.0cm
\caption{Mean-field HDW band structures for (a) $n_c=1/3$ (b) $n_c=3/8$ (c) $n_c=7/16$. The hybridized bands are solid, while the conduction electron dispersion ($J=0$) is shown with a black dashed line. All band structures were obtained self-consistently for $J=8.0$. Despite the changing filling, one of the Fermi surfaces is always pinned at $k_F^*=\pi/2$, since $Q = 4\pi (1/2 - n_c)$. \label{fig:FigS1_mfbs}}
\end{figure}

The self-consistent hybridized band structures are shown for several fillings in the pure HDW region in Fig. \ref{fig:FigS1_mfbs}. Despite the complicated, folded band-structure, one Fermi level crossing is pinned to $\pi/2$ for all fillings, due to the relationship between $n_c$ and $Q$. This crossing at $\pi/2$ is robust for sufficiently large $J$'s ($J \gtrsim 1-2$ for most fillings), although there are small deviations in the self-consistent results for smaller $J$. In addition to the  $\pi/2$ crossing, there can be many other heavy bands crossing the Fermi level at different $k$, however, most do not lead to jumps in the conduction electron momentum distribution, as shown in the main text. Here, we show the heavy ($\beta$) bands, but in DMRG, we can only access the $c$-fermions, which adds a form-factor: 
\begin{align}\label{eq:nqSM}
    n_q& =\frac{1}{L}\sum_{ij\alpha\sigma}e^{iq(i-j)}\langle c\dg_{i\alpha\sigma}c_{j\alpha\sigma}\rangle\cr
    & =\frac{1}{M}\sum_{\gamma\gamma'\alpha\sigma}\langle c\dg_{q\gamma\alpha\sigma}c_{q\gamma'\alpha\sigma}\rangle=\frac{1}{M}\sum_{\eta\gamma\gamma'\alpha\sigma}\langle \beta\dg_{q\eta\sigma}\beta_{q\eta\sigma}\rangle T_{\eta,\gamma\alpha}(q)T^{*}_{\eta,\gamma\alpha}(q),
\end{align}
where $T$ effects the unitary transformation between $\beta$ and $c$, $f$-fermions.
As the mean-field Hamiltonian is diagonal in the $\beta$-fermions, the expectation value above is just the momentum distribution function of $\beta$-fermions:
\begin{equation}
    \langle \beta\dg_{q\eta\sigma}\beta_{q\eta\sigma}\rangle=n_q^{(\beta)}=\theta(-E_{q\eta}),
\end{equation}
leading finally to:
\begin{equation}\label{eq:nqres}
    n_q=\frac{1}{M}\sum_{\eta\gamma\gamma'\alpha\sigma}\theta(-E_{q\eta}) T_{\eta,\gamma\alpha}(q)T^{*}_{\eta,\gamma\alpha}(q).
\end{equation}
The consequence of these $T$ form factors is that most of the Fermi level crossings do not result in $n_q$ jumps, with only the $\pi/2$ and one low-$q$ jump appearing in the large-N mean-field calculation of $n_q$, as shown in Fig. 3(d) in the main text, consistent with the locations of the nonanalyticities found in DMRG.

\section{II. Central charges of HDW phases}

In order to determine the character of critical (gapless) HDW phases, we turn to the calculation of central charges based on cut entanglement entropies. For gapless 1D and quasi-1D systems with open boundaries, the von Neumann entanglement entropy, $S_{vN}$  as a function of cut, $x$ is described by a conformal field theory satisfying the following relation \cite{Vidal2003, Laflorencie2006}: 
\begin{equation}\label{eq:svn_cuts}
    S_{vN}(x)=\frac{c}{6}\log{\left(\frac{2L}{\pi}\sin{ \frac{\pi x}{L}}\right)}+s_0,
\end{equation}
where $c$ is the central charge and $s_0$ is a cut-independent constant contribution.

\begin{figure}[!htb]\vspace*{-0cm}
\includegraphics[width=1.0\columnwidth]{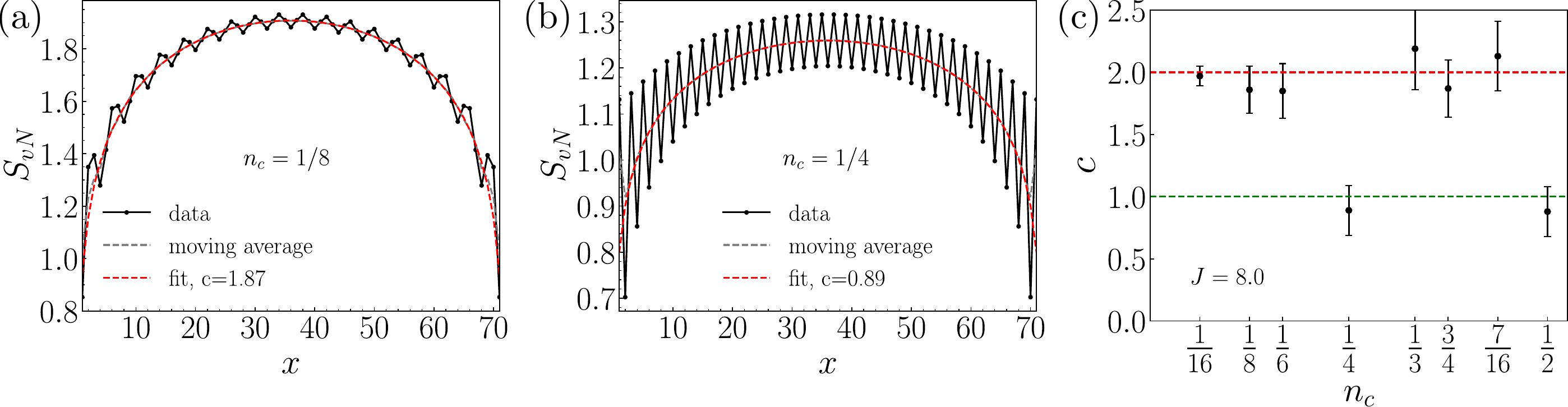}%
\vskip -0.0cm
\caption{Cut entanglement entropy and central charge of HDW phases. (a) For an example critical FM-HDW phase at $n_c=1/4$, $J=8.0$, the cut entanglement entropy follows the behavior of Eq. \ref{eq:svn_cuts}, with additional oscillations induced by the boundaries \cite{Laflorencie2006, Hui-Ke2023}. The fit for the central charge is obtained by taking the moving average of the cut entanglement entropy with a window equal to HDW period, excluding the two boundary windows. The final result of $c=1.87 \pm 0.19$ is consistent with the $c=2$ Tomonaga-Luttinger liquid of charge and channel-spin excitations. (b) The insulating AFH phase at $n_c=1/2$, $J=8.0$ indicates a channel-spin TLL with $c=1$; (c) The critical charge for varying $n_c$ across a  $J=8.0$ phase diagram cut. Both metallic HDW phases are consistent with  a $c=2$ TLL, while the insulating AFH phase at quarter-filling and the insulating AFM phase at half-filling both suggest a $c=1$ TLL. All the results are obtained at $L=72$ and $m=5000$ bond dimension and the value of $c$ does not significantly change further upon increasing the system size and bond dimension.
\label{fig:FigS1b_cs}}
\end{figure}

Representative examples of the entanglement entropy cut dependencies in the HDW phases are shown in Fig. \ref{fig:FigS1b_cs}(a)-(b). The extraction of the central charge is complicated by the appearance of an additional oscillating contribution to the cut entanglement entropy, similar to that observed in \cite{Laflorencie2006, Hui-Ke2023}. These contributions are the consequence of open boundary conditions \cite{Laflorencie2006} and the HDW nature of the phases themselves, as the oscillation period is equal to the HDW period. In order to minimize their effect, we first calculate the moving average of entanglement entropy across a window equal to the oscillation period. The fitting of the central charge according to Eq. \ref{eq:svn_cuts} is then done on the averaged entropy, discarding the first two boundary windows. The fit error is dominated by systematic effects stemming from the oscillations, and we estimate it as a difference between moving average fit (according to the theory in \cite{Laflorencie2006}) and the fit of the lower entanglement entropy peak branch (as employed by \cite{Hui-Ke2023}). We confirm that the value of the fit does not change appreciably upon system size and bond dimension increase.

The fitted results for the central charge are presented in Fig. \ref{fig:FigS1b_cs}(c) for a strong Kondo coupling cut at $J=8$. The central charge is consistent with $c=2$ in the metallic HDW phases, both HDW and FM-HDW, within error bounds, which is consistent with the Tomonaga-Luttinger liquid universality class. The two insulating phases, AFH at quarter-filling and AFM at half-filling, also appear to be TLLs with $c=1$. The difference suggests that the charge sector contributes $c=1$ in the metallic HDW phases, while the channel and spin sectors provide the rest. Note that there is a spin gap at $n_c = 1/4$, which suggests the channel sector provides the full $c = 1$ in the hastatic Kondo insulator. This phase is likely a channel version of the Haldane chain.  The phase at $n_c = 1/2$ is similarly likely a spin version of the Haldane chain, with the spins totally decoupled from the conduction electrons, and so the full $c=1$ here comes from the spin sector.  In the metallic phases, both the spin and channel sectors are gapless, which makes the theoretical understanding of the $c = 2$ central charge in these regions an interesting open question - is there a smooth evolution of the central charge $c=1$ from purely channel to purely spin, or do the metallic regions have Majorana degrees of freedom for the spin/channel ($c = 1 = 1/2 + 1/2$)?

Previous work on the central charges of the two-channel Kondo lattice models, in particular, \cite{Andrei2000, Ge2024}, based on the $SU(2)_2$ Wess-Zumino-Novikov-Witten theory,  has found a fractional central charge, and thus non-TLL critical behavior. However, this work dealt with a multichannel Kondo-Heisenberg model in which the Kondo coupling was small compared to hopping and Heisenberg coupling. This regime, thus, does not apply to the strong Kondo coupling picture in which we find the HDW phases, but rather to weak coupling where we find tentative RKKY liquid signatures. The extraction of central charges in this region cannot be done reliably, as the weak coupling regime is characterized by lower DMRG precision for the largest accessible bond dimensions. We thus leave the question of the nature of the critical weak coupling phase to future work. On the other hand, recent dynamical large-$N$ treatment of the two-channel lattice \cite{Ge2022} has calculated the central charge consistent with $c=1+\gamma$ for incommensurate fillings, where constant $\gamma=4S/N$, with $S$ being the spin and $N$ labeling the $SU(N)$ expansion order. For the  $S=1/2$, $N=2$ limit corresponding to our microscopic model, $c=2$ is obtained, in agreement with our HDW findings.

\section{III. Friedel oscillations and Fermi surfaces}

In the main paper, we used the conduction electron momentum distribution, $n_q$ to determine $k_F^*$ directly.  The charge and spin Friedel oscillations can be used to check these $k_F^*$'s, as well as to obtain information about the charge Luttinger parameter, $K$.

Charge Friedel oscillations are naturally induced by the open boundary conditions \cite{Shibata1996}. Alternatively, we can measure the related charge-charge correlation function, which gives results consistent with charge oscillations.  Spin Friedel oscillations must be artificially induced by applying opposing magnetic fields to the boundaries \cite{Shibata1996}. In all metallic regions explored, both types of Friedel oscillations followed the expected TLL form \cite{Voit1995}, with additional complications arising due to the multiple Fermi wave-vectors in the pure HDW and the small value of $J/t$ in the IC-AFM that weakened the convergence. The simplicity of the FM-HDW Fermi surface made it the most straightforward. In the other regions, the Friedel oscillations were sometimes ambiguous about the Fermi wave-vector, but were still always consistent with the momentum distribution function results.

\begin{figure}[!htb]
\vspace*{-0cm}
\includegraphics[width=0.75\columnwidth]{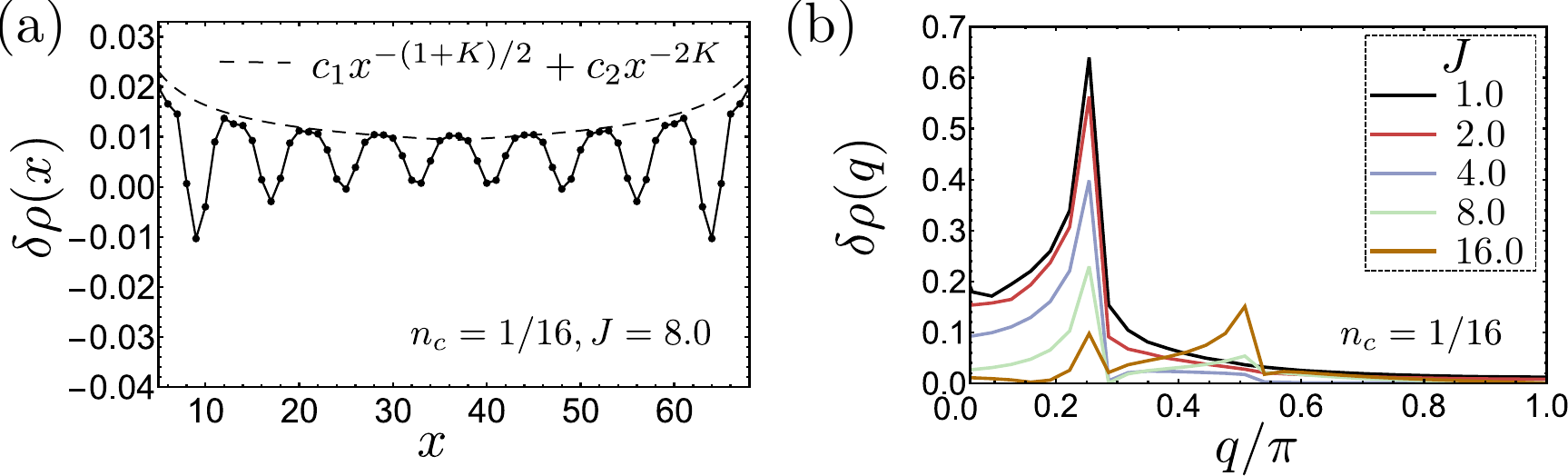}%
\vskip -0.0cm
\caption{(a) Friedel charge oscillations in the FM-HDW showing the behaviour characteristic of a Tomonaga-Luttinger liquid with $k_F^*=2 \pi n_c$ and $K\approx 0.26$, with $c_1=0.044$, $c_2=0.0080$.  (b) Fourier transform of the Friedel charge oscillations for several values of $J$ at $n_c=1/16$. The peaks at $2k_F^*$ and $4k_F^*$ are prominent, with changing intensity due to changing charge $K$ exponent.\label{fig:FigS2_Friedel}}
\end{figure}

Here, we provide an example of Friedel charge oscillations in the FM-HDW, Fig. \ref{fig:FigS2_Friedel}. The charge density operator measured is:
\begin{equation}\label{eq:delrho}
    \delta \rho (x)=\sum_{\alpha\sigma}\langle c\dg_{x\alpha\sigma}c_{x\alpha\sigma} \rangle -4 n_c,
\end{equation}
and the expected TLL form of the oscillations is \cite{Voit1995}:
\begin{equation}\label{eq:Friedel}
     \delta \rho (x)^{TLL}=c_1 \cos{\left(2k_F^{*}x\right)}x^{-(1+K)/2}+c_2 \cos{\left(4k_F^{*}x\right)}x^{-2K},
\end{equation}
where $K$ is the charge Luttinger parameter with $K=K_{\rho}$ for a system without a spin gap (a TLL). From Fig. \ref{fig:FigS2_Friedel} (a),  we see that the calculated oscillations agree well with the TLL form, while from the Fourier transform in Fig. \ref{fig:FigS2_Friedel} (b) we detect the $2k_F^*$ and $4k_F^*$ Fermi wave-vectors of the FM-HDW. The increasing relative weight of the $4k_F^*$ component as $J$ increases implies a decreasing charge Luttinger parameter, $K$, which implies increasingly repulsive interactions between the spinless fermions of the TLL \cite{Voit1995,Shibata1997}.  This feature is more difficult to resolve in the pure HDW, due to the multiple Fermi wave-vectors, but if we focus on the sub-leading low $q$ $k_F^*$, it also appears to be true for those $2k_F^*$ and $4k_F^*$ weights.  The $k_F^* = \pi/2$ leads to degenerate $2k_F^*$ and $4k_F^*$ peaks, making it useless for this analysis. The $K(J)$ dependence found here for the two-channel Kondo model is the opposite of the single-channel case, where $K(J)$ increases as $J$ increases, indicating weakening residual interactions \cite{Shibata1997,Khait2018}. The difference is likely due to the critical nature of overscreened two-channel Kondo model \cite{Cox1998, coleman2015book}, but more exploration is needed to definitively determine the charge Luttinger parameter in the pure HDW, particularly in the light of the multiple Fermi wave-vectors.

\section{IV. Finite-size dependence of hastatic correlations}

\begin{figure}[!htb]
\vspace*{-0cm}
\includegraphics[width=1.0\columnwidth]{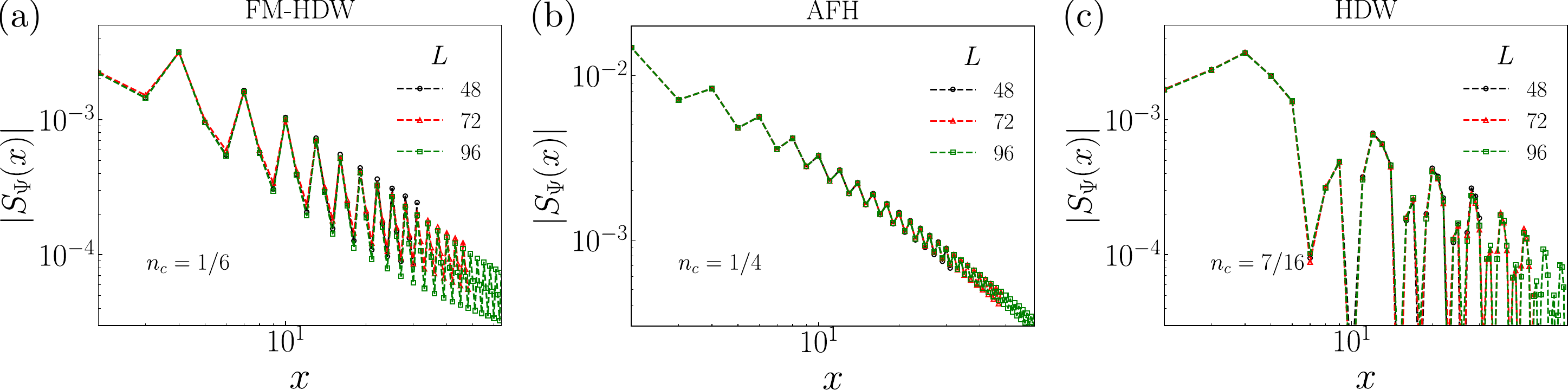}%
\vskip -0.0cm
\caption{Hastatic order parameter correlations as a function of increasing system size. No change is observed in the algebraic decay of hastatic correlation upon system size increase from 48 to 96 sites, as exemplified by (a) FM-HDW (b) AFH-HDW (c) HDW examples. The results shown are for $J=8.0$ and the maximum bond dimension $m=5000$. Small deviations at large $x$ are the consequence of approaching the second boundary of the system.
\label{fig:FigS2b_Lscale}}
\end{figure}

In order to definitively establish the algebraic decay of hastatic correlations, we have been particularly careful to exclude finite system size and finite bond dimension effects up to the limits of our numerics. As a typical example, Fig. \ref{fig:FigS2b_Lscale} show hastatic correlations at an illustrative point in each of the HDW phases as a function of increasing system size. There is no significant change in the algebraic decay of oscillations upon increasing system size up to $L=96$; the minor differences seen at large $x$ are the consequence of approaching the second boundary of the system at smaller system sizes.

\section{V. Composite pairing correlations}

As composite pair superconductivity has long been proposed in two-channel Kondo lattices \cite{Coleman1999,Flint2008, Hoshino2014}, we explicitly looked for composite pair correlations throughout our phase diagram.  Tomonaga-Luttinger liquids naturally have  algebraic spin singlet and spin triplet superconducting correlations \cite{Voit1995}; these composite pair correlations involve the spins specifically. As a reminder, the composite pair order parameter is, 
\begin{equation}\label{eq:CPOP_sm}
    \Delta_{CP}(j)=\sum_{\alpha\sigma\sigma'}c\dg_{j\alpha\sigma'}\left[\vect{\sigma} (i\sigma_2)\right]_{\sigma\sigma'}c\dg_{j\bar{\alpha}\sigma'}\cdot \vect{S}_{fj},
\end{equation}
where $\bar{\alpha}$ is the conduction electron channel orthogonal to $\alpha$. This superconducting pair explicitly includes electrons from both channels, as well as the local spin.  To explore the susceptibility to composite pairing, we calculated the corresponding real space composite pair correlations, given by:
\begin{equation}\label{eq:SOP_sm}
    S_{CP}(x)=\langle \Delta_{CP}(j) \Delta\dg_{CP}(j+x)\rangle,
\end{equation}
where, as in the main text, we fixed $j = 10$, which gives generic results.

\begin{figure}[!htb]
\vspace*{-0cm}
\includegraphics[width=1.0\columnwidth]{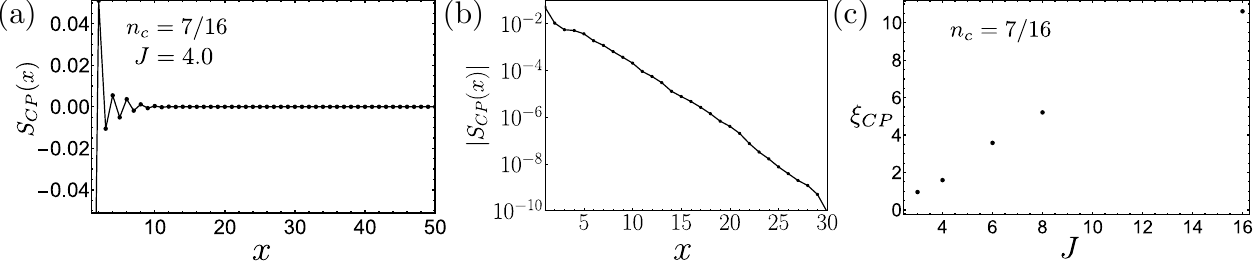}%
\vskip -0.0cm
\caption{(a) Composite pair correlations, $S_{CP}(x)$ for  $J=4.0$ and $n_c=7/16$. The correlations decay exponential and are staggered. (b) The semilog plot of the composite pair correlations from (a), confirming the exponential decay. (c) Composite pair correlation lengths, $\xi_{CP}$ as a function of $J$ for $n_c=7/16$. While composite pairing correlations decay exponentially everywhere in the phase diagram, the correlation length in the HDW grows linearly with $J$.\label{fig:FigS3_CP}}
\end{figure}

In Fig. \ref{fig:FigS3_CP}(a)-(b), we show a typical example in the HDW, where the correlations clearly decay exponentially, with a modulation corresponding to staggered composite pairs.  The decay is particularly rapid in the FM-HDW and IC-AFM, where correlations do not extend beyond the initial site. In the pure HDW, the correlation length is larger, and can be determined from the expected $S_{CP}(x)=c_0 e^{-x/\xi_{CP}}$ form.  $\xi_{CP}$ increases with increasing $J$, as shown in  Fig. \ref{fig:FigS3_CP}(c). The increase is roughly linear in $J$ (only in the HDW), which suggests composite pair superconductivity might be competitive in the $J \rightarrow \infty$ limit and might indicate that composite superconductivity is sometimes the ground state in higher dimensions.

The observed staggered composite pair correlations are reminiscent of the staggered composite superconductivity found in $d = \infty$ for the two-channel lattice \cite{Hoshino2014}. The region of significant $\xi_{CP}$ ($1/4<n_c<1/2$, pure HDW) corresponds well to the fillings at which the staggered composite superconductivity appears in $d = \infty$, and $\xi_{CP}$ increases with both $J$, as shown, and $n_c$, qualitatively similar to the infinite dimensional behaviour. Future DMRG work in two-dimensions could provide interesting insights into the competition between composite superconductivity and hastatic order.

\section{VI. DMRG details and reference values for the ground state energy}

 In order to obtain the ground state phase diagram, we employed finite system DMRG \cite{White1992,White1993} in an ITensor implementation \cite{itensor} with open boundary conditions.  We conserved the total conduction electron number, $n_c=\frac{1}{4L}\sum_{i\sigma\alpha}n_{i\alpha\sigma}$,
 and the $z$ component of total angular momentum, $S^{z}=\sum_{i}\left(S_{fi}^z+\sum_{\alpha}S_{ci\alpha}^z\right)$, where $\vect{S}_{ci\alpha}=\frac{1}{2}\sum_{\sigma\sigma'}c\dg_{i\alpha\sigma}\vect{\sigma}_{\sigma\sigma'}c_{i\alpha\sigma'}$ is the conduction electron spin for a given site and channel. We used bond dimensions of up to $m=5000$ on lattices of up to $L=96$ sites. In order to facilitate convergence and avoid local minima, we did a significant number of small bond-dimension sweeps with noise term included (starting from 10$^{-6}$). The bond dimension was then progressively increased with the decreasing noise term to 0 around $m=2000$. This resulted in a maximum discarded weight of $<10^{-6}$ across the whole phase diagram. The largest discarded weights were seen in the weak coupling regime, while for $J>t$, the maximum discarded weight observed was $<10^{-8}$. This further decreased with the increasing $J/t$ ratio, particularly in the insulating phases (at quarter- and half-filling) where the discarded weights of $<10^{-11}$  are typical. Convergence was further ensured by the expectation values of relevant physical quantities - energies, total angular momenta, correlators - being not affected upon the largest bond dimension increase (at most $10^{-5}$ relative change in energy at the largest bond dimension increase for $J>t$). All figures were produced with $L = 72$, unless otherwise specified, although convergence with increasing system size was explicitly explored, as described in Sec. IV.

Here we show several reference values for the ground state energy obtained by the DMRG in Table. \ref{tab:EGS}.
\begin{table}[htb]
\begin{tabular}{@{}ccccc@{}}
\toprule[1pt]
$n_c$ && $J$ &&   $E_{GS}/L$ \\\midrule[0.75pt]
1/2  && 8   && -8.65967889  \\
7/16 && 8   && -8.29215291  \\
5/12 && 8   && -8.13989240  \\
3/8  && 8   && -7.79078002  \\
1/3  && 8   && -7.39022144  \\
1/4  && 8   && -6.46129200  \\
1/4  && 16  && -12.25758904 \\
1/4  && 4   && -3.71723236  \\
1/4  && 2   && -2.45941501  \\
1/4  && 1   && -1.90225924  \\
1/4  && 1/2 && -1.74500862 \\ \bottomrule[1pt]
\end{tabular}
\caption{Reference values of the DMRG obtained the ground state energy per site for several different conduction electron fillings and Kondo couplings. The results shown are for $L=72$.\label{tab:EGS}}
\end{table}

\end{document}